\documentclass[full paper,twocolumn]{jpsj2}

\usepackage{graphicx}

\def\ve#1{\mbox{\boldmath $#1$}}

\def\Tc{T_\mathrm{c}}
\def\Re{\mathrm{Re}}

\title{%
Perturbation Analysis of Superconductivity in the Trellis-Lattice Hubbard Model \\
}

\author{%
Sotaro \textsc{Sasaki}
\thanks{E-mail address: sotaro@scphys.kyoto-u.ac.jp},
Hiroaki \textsc{Ikeda}
and Kosaku \textsc{Yamada}
}

\inst{
Department of Physics, Kyoto University, Sakyo-ku, Kyoto 606-8502 \\
}

\recdate{\today}

\abst{%
We investigate pairing symmetry and transition temperature 
in the trellis-lattice Hubbard model. 
We solve the \'Eliashberg equation using the third-order perturbation 
theory with respect to the on-site repulsion $U$. 
We find that a spin-singlet state is very stable in a wide range of 
parameters. 
On the other hand, when the electron number 
density is shifted from the half-filled state and the band gap 
between two bands 
is small, a spin-triplet superconductivity is expected. 
Finally, we discuss a possibility of unconventional superconductivity 
and pairing symmetry in Sr$_{14-x}$Ca$_x$Cu$_{24}$O$_{41}$. 
}

\kword{%
trellis lattice, quasi-one-dimensional superconductors, pairing symmetry, 
transition temperature, third-order perturbation theory
}

\begin{document}
\sloppy
\maketitle

\section{Introduction}
Quasi-one-dimensional superconductors have been studied 
and attracted our attention. 
Today, some quasi-one-dimensional superconductors, 
such as (TMTSF)$_2$X~\cite{rf:Ishiguro,rf:Jerome} 
and $\beta$-Na$_{0.33}$V$_2$O$_5$~\cite{rf:Yamauchi}, were discovered, 
and their superconductivity has been investigated. 
In 1996, a superconducting transition in Sr$_{14-x}$Ca$_x$Cu$_{24}$O$_{41}$ 
was discovered at the transition temperature $\Tc\simeq 12\,$K under 
high pressure of approximately 3$\,$GPa for $x=13.6$.~\cite{rf:Uehara} 
This material possesses a quasi-one-dimensional lattice structure 
called trellis lattice, 
which is Cu network connected by O orbitals similar to high-$\Tc$ 
superconductors as shown in Fig. \ref{fig:lattice}(a). 
Actually, this material shows quasi-one-dimensional metallic behavior 
in an electric resistivity experiment.\cite{rf:Nagata} 
The ratio of the resistivities $\rho_a/\rho_c$ is approximately 80 
at ambient pressure, and is reduced approximately 30 under 3.5 GPa 
at $T\simeq 50$$\,$K 
for $x=11.5$. 
Here, the Cu valence in this ladder can be changed from $+2.07$ ($x=0$) 
to $+2.24$ 
($x=14$) by substituting Ca for Sr~\cite{rf:Osafune}. 

Recently, an NMR experiment has been performed in this material for 
$x=12$ by Fujiwara {\it et al}.
~\cite{rf:Fujiwara}  
From this experiment, an activated $T$ dependence of $1/T_1$ is observed 
at temperatures higher than $30\,$K, 
and it suggests that the spin gap is conformed. 
Below $30\,$K, $1/T_1T$ keeps constant and shows the Fermi liquid behavior. 
Moreover, $1/T_1$ has a small peak just below $\Tc$. 
It indicates that the superconducting gap structure is a fully gapped 
state in this material. 
On the other hand, the Knight shift does not change above and below $\Tc$. 
It suggests that a spin-triplet state is realized. But, since the paramagnetic 
contribution of Knight shift should be small owing to the effect of spin 
gap conformed at rather high temperatures, it might be 
difficult to detect the shift at $\Tc$ within the experimental accuracy. 

The theoretical investigations for the possibility of unconventional 
superconductivity 
have been reported in the quasi-one-dimensional superconductors, such as 
(TMTSF)$_2$X, $\beta$-Na$_{0.33}$V$_2$O$_5$ and 
Sr$_{14-x}$Ca$_x$Cu$_{24}$O$_{41}$. 
The superconductivity in (TMTSF)$_2$X has been investigated within the 
fluctuation-exchange approximation (FLEX)~\cite{rf:Kino} and 
the third-order perturbation theory (TOPT)~\cite{rf:Nomura}. 
The superconductivity in $\beta$-Na$_{0.33}$V$_2$O$_5$ has been 
investigated within TOPT.~\cite{rf:Sasaki} 
Also, on the basis of the calculation within FLEX for the 
trellis-lattice Hubbard model, Kontani and Ueda indicated that a spin-singlet 
and fully gapped 
state is stable in Sr$_{14-x}$Ca$_x$Cu$_{24}$O$_{41}$.~\cite{rf:Kontani} 
In this paper, by using the third-order perturbation theory~\cite{rf:Yanase}, 
we investigate 
in detail unconventional superconductivity in 
the trellis-lattice Hubbard model, keeping Sr$_{14-x}$Ca$_x$Cu$_{24}$O$_{41}$ 
in mind. In particular, we point out that the fully gapped state can be 
realized for unconventional superconductivity, and the spin-triplet state can 
be realized in a range of parameters. 

\section{Model}
\begin{figure}[t]
\begin{center}
\includegraphics[width=8cm]{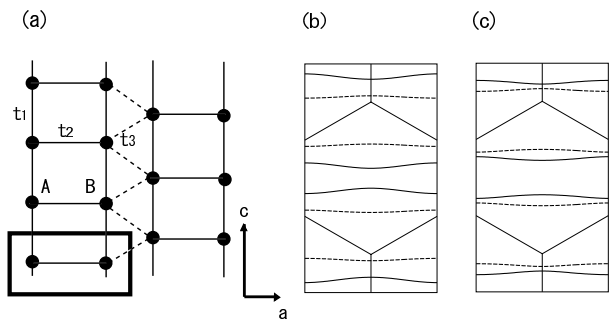}
\end{center}
\caption{(a) Schematic figure of the lattice used in this calculation. 
$t_i(1\leq i \leq 3)$ is the hopping integral. 
The region enclosed by a rectangle is a primitive 
cell. The primitive cell topologycally composes a triangular lattice. 
(b) The Fermi surfaces for electron number density per ladder site $n=0.80$ 
and the hopping integrals $t_2=1.0$, $t_3=0.15$ 
are shown by the solid lines for $\varepsilon(k)+V(k)$ and by the dashed line 
for $\varepsilon(k)-V(k)$, respectively.  
Since the lattice is topologycally triangular lattice, 
the Brillouin zone is hexagonal.
(c) The Fermi surface for $n=0.80$, $t_1=1.0$, $t_2=0.40$, $t_3=0.10$. 
In this case, the band gap is small. }
\label{fig:lattice}
\end{figure}
Now, let us consider the lattice structure and the band structure in 
Sr$_{14-x}$Ca$_x$Cu$_{24}$O$_{41}$.
We can consider the lattice structure with 
the Cu network called trellis lattice shown 
in Fig. \ref{fig:lattice}(a). 
The unit cell is a thick-line rectangle, and contains two sites A and B. 
We use a simple tight-binding model. 
In this case, we consider three types of hopping integrals 
displayed in Fig. \ref{fig:lattice}(a).

Here, we investigate in detail the nature of superconductivity in 
such a situation. 
We consider the quasi-one-dimensional two band repulsive Hubbard model. 
\begin{equation}
\begin{split}
H&=H_0+H_{\rm int},
\end{split}
\end{equation}
where
\begin{equation}
\begin{split}
H_0&=\sum_{k,\sigma}\varepsilon(k)A^{\dagger}_{k\sigma}
A_{k\sigma}+\sum_{k,\sigma}\varepsilon(k)B^{\dagger}_{k\sigma}B_{k\sigma}\\
&+\sum_{k,\sigma}V(k)A^{\dagger}_{k\sigma}B_{k\sigma}
+\sum_{k,\sigma}V(k)B^{\dagger}_{k\sigma}A_{k\sigma},
\end{split}
\end{equation}
~\cite{rf:comment}
and
\begin{equation}
\begin{split}
H_{\rm int}&=\frac{U}{2N}\sum_{k_i}\sum_{\sigma\neq\sigma'}
A^{\dagger}_{k_1\sigma}A^{\dagger}_{k_2\sigma'}A_{k_3\sigma'}
A_{k_4\sigma}\delta_{k_1+k_2,k_3+k_4}\\
&+\frac{U}{2N}\sum_{k_i}\sum_{\sigma\neq\sigma'}
B^{\dagger}_{k_1\sigma}B^{\dagger}_{k_2\sigma'}B_{k_3\sigma'}
B_{k_4\sigma}\delta_{k_1+k_2,k_3+k_4}.
\end{split}
\end{equation}
From the tight binding approximation, 
\begin{equation}
\begin{split}
&\varepsilon(k)=-2t_1\cos(k_c), \\
&V(k)=|-t_2-2t_3\cos(k_c/2)e^{i\sqrt{3}k_a/2}|. 
\end{split}
\end{equation}
Here, $A_{k,\sigma}$ and $B_{k,\sigma}$($A_{k,\sigma}^{\dagger}$ and  
$B_{k,\sigma}^{\dagger})$ are the annihilation (creation) operator for the 
electron at A and B site, respectively. 
To obtain two bands, we transform $A_{k,\sigma}^{({\dagger})}$, 
$B_{k,\sigma}^{({\dagger})}$ into $a_k^{({\dagger})},b_k^{({\dagger})}$ as
\begin{equation}
\begin{split}
a_k^{({\dagger})}&=\frac{1}{\sqrt{2}}A_k^{({\dagger})}+\frac{1}
{\sqrt{2}}B_k^{({\dagger})}, \\
b_k^{({\dagger})}&=-\frac{1}{\sqrt{2}}A_k^{({\dagger})}+\frac{1}
{\sqrt{2}}B_k^{({\dagger})} .\label{eq:trans}
\end{split}
\end{equation}
With this transformation, $H_0$ is transformed into 
\begin{equation}
\begin{split}
H_0=\sum_{k,\sigma}(\varepsilon(k)+V(k))a^{\dagger}_{k\sigma}
a_{k\sigma}+\sum_{k,\sigma}(\varepsilon(k)-V(k))
b^{\dagger}_{k\sigma}b_{k\sigma}. 
\end{split}
\end{equation}
Thus, we obtain two bands, $\varepsilon(k)+V(k)$ and $\varepsilon(k)-V(k)$. 
We use $t_1=1.0$ as an energy unit. 
In the previous calculation within FLEX~\cite{rf:Kontani}, $t_2=1.0$ and $t_3=0.15$ are used to fit the band structure calculated within the local-density 
approximation~\cite{rf:Arai}. 
In the calculation, 
the electron number density per ladder site $n=1.0$ is assumed. 
It means that the valence of Cu is $+2.0$. 
On the other hand, we use the hopping integrals $t_2$, $t_3$ and 
electron number density per ladder site $n$ as a parameter in this 
calculation. $t_2$ is a measure of the band gap between two bands, 
and $t_3$ is related to one-dimensionality. 
We show typical quasi-one-dimensional Fermi surfaces 
in Fig. \ref{fig:lattice}(b) and (c). 
We show the Fermi surface for $t_2=1.0$, $t_3=0.15$, $n=0.80$ 
in Fig. \ref{fig:lattice}(b), 
and for $t_2=0.40$, $t_3=0.10$, $n=0.80$ in Fig. \ref{fig:lattice}(c), 
respectively. 
Actually, the band gap is small in Fig. \ref{fig:lattice}(c). 

\section{Formulation}
\begin{figure}[t]
\begin{center}
\includegraphics[width=8cm]{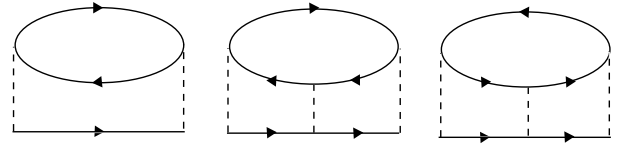}
\end{center}
\caption{The diagrams in the normal self-energy. 
The solid and the broken lines 
represent $G^{(0)}_{\alpha\beta}(k)$ and $U$, 
respectively. 
}
\label{fig:selfenergy}
\end{figure}

We apply the third-order perturbation theory 
with respect to $U$ to our model. 
The diagrams in the normal self-energy are 
shown in Fig. \ref{fig:selfenergy}. 
The normal self-energy is given by
\begin{equation}
\begin{split}
&\Sigma^{\rm N} _{\alpha\beta}(k)=\frac{T}{N}\sum_{k',\gamma} [\chi^{(0)}_{\alpha\beta}(k-k') G^{(0)}_{\alpha\beta}(k')U^2+ \\
&( \chi^{(0)}_{\alpha\gamma}(k-k')\chi^{(0)}_{\gamma\beta}(k-k')
+\phi^{(0)}_{\alpha\gamma}(k+k')\phi^{(0)}_{\gamma\beta}(k+k')) \\
&\times G^{(0)}_{\alpha\beta}(k')U^3], 
\end{split}
\end{equation}
where 
\begin{equation}
\begin{split}
&G^{(0)}_{\alpha\beta}(k)=\frac{1}{2}(G^{(0)}_{+}(k)+G^{(0)}_{-}(k))\quad (\alpha=\beta), \\
&G^{(0)}_{\alpha\beta}(k)=\frac{1}{2}(G^{(0)}_{+}(k)-G^{(0)}_{-}(k))\quad (\alpha\neq\beta),\\
&G^{(0)}_{\pm}(k)=\frac{1}{i\omega_n-(\varepsilon (\ve{k})\pm V(\ve{k}))+\mu}, \\
&\chi^{(0)}_{\alpha\beta}(q)=-\frac{T}{N}\sum_{k} G^{(0)}_{\alpha\beta}(k)G^{(0)}_{\beta\alpha}(q+k), \\
&\phi^{(0)}_{\alpha\beta}(q)=-\frac{T}{N}\sum_{k}G^{(0)}_{\alpha\beta}(k)G^{(0)}_{\alpha\beta}(q-k).
\end{split}
\end{equation}
$G^{(0)}_{\alpha \beta}(k)$ with the short notation 
$k=(\ve{k},\omega_n)$ represents the bare Green's function. 
$\alpha$ and $\beta$ represent A or B. 
Here, 
\begin{equation}
\begin{split}
&\Sigma^{\rm N} _{\rm AA}(k)=\Sigma^{\rm N} _{\rm BB}(k),\\ 
&\Sigma^{\rm N} _{\rm AB}(k)=\Sigma^{\rm N} _{\rm BA}(k).
\end{split}
\end{equation} 
Since the first-order normal self-energy is constant, it can be included by 
the chemical potential $\mu$.
The dressed Green's function $G_{\alpha\beta}(k)$ is given by
\begin{equation}
\begin{split}
&G_{\alpha\beta}(k)=\frac{1}{2}(G_{+}(k)+G_{-}(k))\quad (\alpha=\beta),\\
&G_{\alpha\beta}(k)=\frac{1}{2}(G_{+}(k)-G_{-}(k))\quad (\alpha\neq\beta),\\
&G_{\pm}(k)=\frac{1}{(i\omega_n-(\varepsilon(\ve{k})\pm V(\ve{k}))-\Sigma^{\rm N}_{\pm}(k)+\mu+\delta\mu)}, \\
&\Sigma^{\rm N}_{\pm}(k)=\Sigma^{\rm N}_{\rm AA}(k)\pm\Sigma^{\rm N}_{\rm AB}(k).
\end{split}
\end{equation}
Here, the chemical potential $\mu$ and the chemical potential shift 
$\delta\mu$ are determined so as to fix the electron number density 
per ladder site $n$, 
\begin{equation}
\begin{split}
n=\frac{T}{N}\sum_{k,\gamma} G^{(0)}_{\gamma\gamma}(k)
=\frac{T}{N}\sum_{k,\gamma} G_{\gamma\gamma}(k).
\end{split}
\end{equation}

\begin{figure}[t]
\begin{center}
\includegraphics[width=8cm]{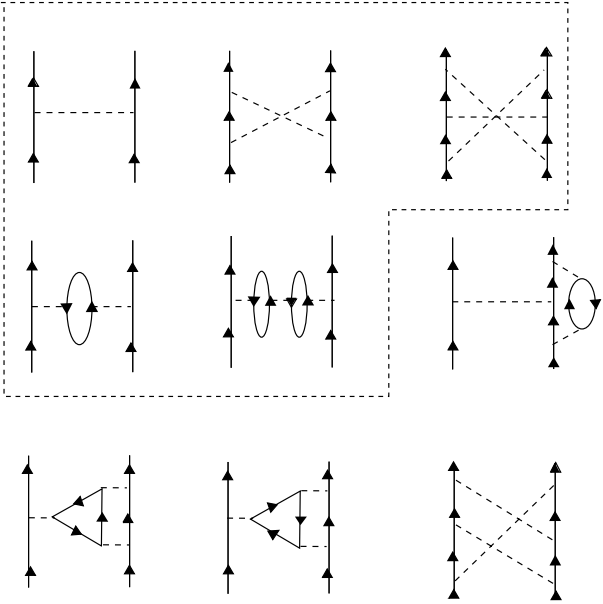}
\end{center}
\caption{The diagrams of pairing interaction part. The diagrams enclosed by 
the dashed line are called the RPA-like terms. The other diagrams are the vertex corrections. }
\label{fig:vertex}
\end{figure}

We also expand the effective pairing interaction up to the third order 
with respect to $U$. 
The diagrams of effective pairing interaction are shown 
in Fig. \ref{fig:vertex}.
For the spin-singlet state, the effective pairing interaction is given by
\begin{equation}
\begin{split}
&V^{\rm Singlet}_{\alpha\beta,\alpha'\beta'}(k;k')\\
&=V^{\rm RPA\,Singlet}_{\alpha\beta,\alpha\beta}(k;k')
+V^{\rm Vertex\,Singlet}_{\alpha\beta,\alpha\gamma}(k;k'),
\end{split}
\end{equation}
where
\begin{equation}
\begin{split}
&V^{\rm RPA\,Singlet}_{\alpha\beta,\alpha\beta}(k;k')=U\delta_{\alpha\beta}
+U^2\chi^{(0)}_{\alpha\beta}(k-k')\\
&+2U^3\chi^{(0)}_{\alpha\gamma}(k-k')\chi^{(0)}_{\gamma\beta}(k-k'), 
\end{split}
\end{equation}
and
\begin{equation}
\begin{split}
&V^{\rm Vertex\,Singlet}_{\alpha\beta,\alpha\gamma}(k;k')=2(T/N)\Re\Big [\sum_{k_1}G^{(0)}_{\beta\alpha}(k_1) \\
&\times(\chi^{(0)}_{\beta\gamma}(k+k_1)-\phi^{(0)}_{\beta\gamma}(k+k_1))G^{(0)}_{\alpha\gamma}(k+k_1-k')U^3\Big ].
\end{split}
\end{equation}

For the spin-triplet state,
\begin{equation}
\begin{split}
&V^{\rm Triplet}_{\alpha\beta,\alpha'\beta'}(k;k')\\
&=V^{\rm RPA\,Triplet}_{\alpha\beta,\alpha\beta}(k;k')
+V^{\rm Vertex\,Triplet}_{\alpha\beta,\alpha\gamma}(k;k'), 
\end{split}
\end{equation}
where
\begin{equation}
\begin{split}
V^{\rm RPA\,Triplet}_{\alpha\beta,\alpha\beta}(k;k')=-U^2\chi^{(0)}_{\alpha\beta}(k-k'), 
\end{split}
\end{equation}
and
\begin{equation}
\begin{split}
&V^{\rm Vertex\,Triplet}_{\alpha\beta,\alpha\gamma}(k;k')=2(T/N)\Re\Big [\sum_{k_1}G^{(0)}_{\beta\alpha}(k_1)\\
&\times (\chi^{(0)}_{\beta\gamma}(k+k_1)+\phi^{(0)}_{\beta\gamma}(k+k_1))G^{(0)}_{\alpha\gamma}(k+k_1-k')U^3\Big ].
\end{split}
\end{equation}
Here, $V^{\rm RPA\,Singlet\,(Triplet)}_{\alpha\beta,\alpha\beta}(k,k')$ is called the RPA-like terms and 
$V^{\rm Vertex\,Singlet\,(Triplet)}_{\alpha\beta,\alpha\gamma}(k,k')$ is called the vertex 
corrections. 
Near the transition point, the anomalous self-energy $\Delta_{\alpha\beta}(k)$ satisfies 
the linearized \'Eliashberg equation,
\begin{equation}
\begin{split}
&\lambda_{\rm max}\Delta_{\alpha\beta}(k)\\
&=-\frac{T}{N}\sum_{k',\gamma}V_{\alpha\beta,\alpha'\beta'}(k;k')F_{\alpha'\beta'}(k'),\\
&=-\frac{T}{N}\sum_{k',\gamma,\alpha'',\beta''}V_{\alpha\beta,\alpha'\beta'}(k;k')G_{\alpha'\alpha''}(k')G_{\beta'\beta''}(-k')\Delta_{\alpha''\beta''}(k')
\end{split}
\end{equation}
~\cite{rf:comment2}
where, $V_{\alpha\beta,\alpha'\beta'}(k;k')$ is $V^{\rm Singlet}_{\alpha\beta,\alpha'\beta'}(k;k')$ or $V^{\rm Triplet}_{\alpha\beta,\alpha'\beta'}(k;k')$. 
Here, $F_{\alpha'\beta'}(k')$ is anomalous Green's function, 
and $\lambda_{\rm max}$ is the largest positive eigenvalue. 
Then, the temperature at $\lambda_{\rm max}=1$ corresponds to $T_{\rm c}$.
By estimating $\lambda_{\rm max}$, 
we can determine which type of pairing symmetry is stable. 
For numerical calculations, we take 128 $\times$ 128 $\ve{k}$-meshes 
for twice space of the first Brillouin zone and 2048 Matsubara frequencies.

\section{Numerical results}
\label{section:res}
\subsection{Dependence of $\lambda_{\rm max}$ on the parameters $t_2$ and 
$t_3$}
\label{subsection:t2t3}
\begin{figure}[t]
\begin{center}
\includegraphics[width=8cm]{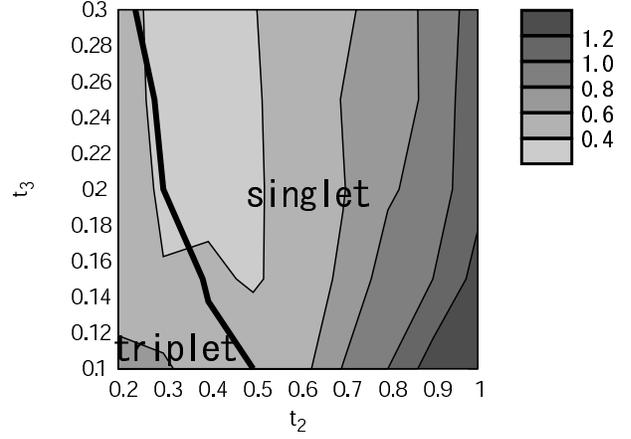}
\end{center}
\caption{A contour plot of $\lambda_{\rm max}$
as a function of $t_2$ and $t_3$ in the case of $T=0.01$, $n=0.80$ and 
$U=4.0$. 
In the right hand side region of the thick line, a spin-singlet state 
is stable, and 
in the left hand side region of the thick line, a spin-triplet state 
is stable. 
} 
\label{fig:t2t3}
\end{figure}
Fig. \ref{fig:t2t3} is a contour plot of $\lambda_{\rm max}$ 
as a function of 
$t_2$ and $t_3$ in the case of $T=0.01$, $n=0.80$ and $U=4.0$. 
In the dark region, the state is stable. 
In the right hand side region of the thick line, a spin-singlet state 
is stable, and 
in the left hand side region of the thick line, a spin-triplet state 
is stable. 
When $t_2$ is large, a spin-singlet state is very stable. 
On the other hand, when $t_2$ is small, a spin-triplet state is stable. 
Therefore, a spin-triplet state is stable when the band gap between two band 
is small. 
We discuss this result later in Sec. \ref{section:dis}. 
For $t_3 < 0.1$, 
the mass enhancement factor is much smaller than unity.
Therefore, reliable numerical calculations in this framework can not be 
obtained in the range 
of $t_3 < 0.1$. 
\subsection{Temperature dependence}
\begin{figure}[t]
\begin{center}
\includegraphics[width=8cm]{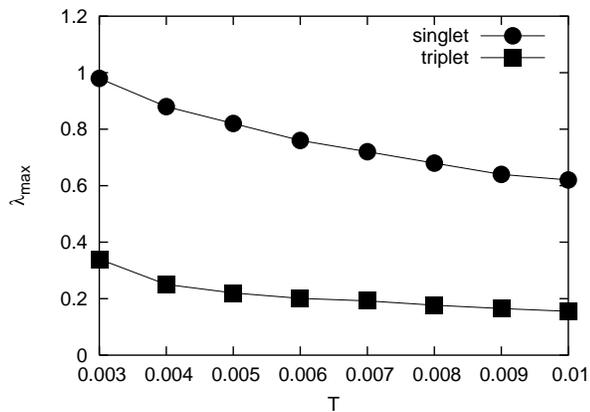}
\end{center}
\caption{Temperature dependence of $\lambda_{\rm max}$ for 
a spin-singlet 
(or spin-triplet) state in the case of 
$t_2=1.0$, $t_3=0.15$, $n=0.80$ and $U=3.2$. 
The line with black circles (squares) is the 
temperature dependence 
for the spin-singlet (spin-triplet) state obtained using the third-order 
perturbation theory. 
The spin-singlet state is more stable than the spin-triplet state in this 
case.}
\label{fig:temp1032}
\end{figure}
\begin{figure}[t]
\begin{center}
\includegraphics[width=8cm]{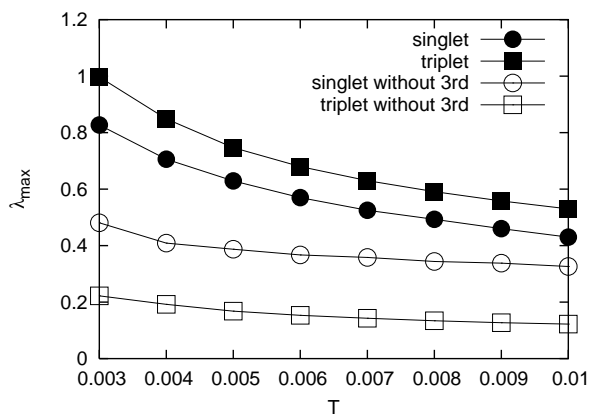}
\end{center}
\caption{Temperature dependence of $\lambda_{\rm max}$ for spin-singlet 
(or spin-triplet) state in the case of $t_2=0.40$, 
$t_3=0.10$, $n=0.80$ and $U=4.1$. 
The line with black circles (square) is the result 
for the spin-singlet (spin-triplet) state obtained using the third-order 
perturbation theory. 
The spin-triplet state is more stable than the spin-singlet state in this 
case. 
The line with the white circles (square) is the result 
for spin-singlet (spin-triplet) state without the third-order terms 
of the pairing interaction. }
\label{fig:temp0441}
\end{figure}
In Fig. \ref{fig:temp1032}, we show temperature dependences of 
$\lambda_{\rm max}$ 
in the case with $t_2=1.0$, $t_3=0.15$ $n=0.80$ and $U=3.2$. 
Also, in Fig. \ref{fig:temp0441}, we show temperature dependences for 
$\lambda_{\rm max}$ 
in the case with $t_2=0.40$, $t_3=0.10$, $n=0.80$ and $U=4.1$. 
With decreasing temperature, $\lambda_{\rm max}$ increases. 
In Fig. \ref{fig:temp1032}, the spin-singlet state is stable. 
On the other hand, in Fig. \ref{fig:temp0441}, the spin-triplet state is 
stable.
These states possess almost the same transition temperature, 
$\Tc\simeq 0.003$, respectively. 
If we assume that the bandwidth $W \sim 6$ corresponds to $2\,{\rm eV}$, 
then $\Tc\sim 10\,{\rm K}$ is obtained in almost accordance with 
the experimental value for Sr$_{14-x}$Ca$_x$Cu$_{24}$O$_{41}$.
In Fig. \ref{fig:temp0441}, we also show the results for $\lambda_{\rm max}$ 
obtained without the pairing interaction due to the third-order terms. 
We can see that the vertex corrections are 
important for stabilizing the spin-triplet state from the comparison. 

\subsection{Dependence of $\lambda_{\rm max}$ on the parameters $t_2$ and $n$}
\begin{figure}[t]
\begin{center}
\includegraphics[width=8cm]{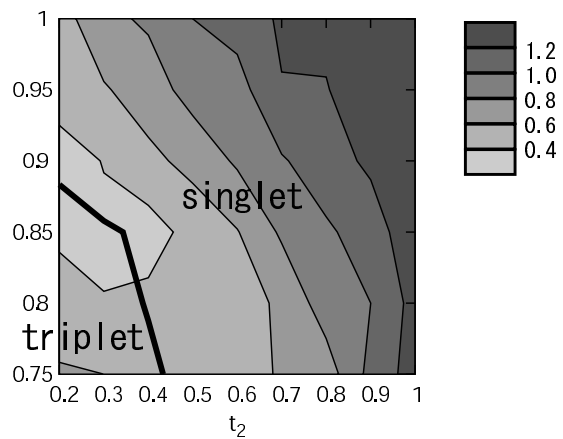}
\end{center}
\caption{
A contour plot of $\lambda_{\rm max}$
as a function of $t_2$ and $n$ in the case of $t_3=0.15$, $T=0.01$ 
and $U=4.0$. 
In the right hand side region of the thick line, a spin-singlet state 
is stable, and 
in the left hand side region of the thick line, a spin-triplet state 
is stable. }
\label{fig:t2n}
\end{figure}
Fig. \ref{fig:t2n} is a contour plot of $\lambda_{\rm max}$ as a function of
$t_2$ and $n$ in the case of $t_3=0.15$, $T=0.01$ and $U=4.0$. 
When $t_2$ is large, a spin-singlet state is very stable like the above 
case in Sec. \ref{subsection:t2t3}. 
On the other hand, if $t_2$ is small and $n$ is shifted from the half-filled 
state, a spin-triplet state is stable. Thus, the spin-triplet state is 
suppressed in the vicinity of the half-filled state, and it is stable rather 
far from the half-filled state. This tendency is a general property. 
The pairing interaction important for the spin-triplet state originates from 
the third-order terms which vanishes in the case with the particle-hole 
symmetry. 
Therefore, $\lambda_{\rm max}$ is reduced due to approximate particle-hole 
symmetry near the half-filled state. 
\subsection{Dependence of $\lambda_{\rm max}$ on the parameters $t_2$ and $U$}
\begin{figure}[t]
\begin{center}
\includegraphics[width=8cm]{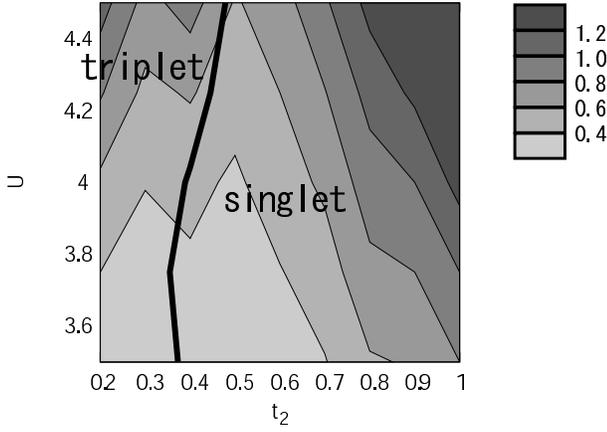}
\end{center}
\caption{A contour plot of $\lambda_{\rm max}$
as a function of $t_2$ and $n$ in the case of $t_3=0.15$, $T=0.01$, 
and $U=4.0$. 
In the right hand side region of the thick line, a spin-singlet state 
is stable, and 
in the left hand side region of the thick line, a spin-triplet state 
is stable.}
\label{fig:t2u}
\end{figure}
Fig. \ref{fig:t2u}, is a contour plot of $\lambda_{\rm max}$ as a function of 
$t_2$ and $U$ in the case of $t_3=0.15$, $T=0.01$ and $n=0.80$. 
When $t_2$ is large, a spin-singlet state is very stable like the above 
case in Sec. \ref{subsection:t2t3}. 
On the other hand, when $U$ becomes large, $\lambda_{\rm max}$ becomes large. 
But the stable symmetry does not depend on the magnitude of $U$ so much in 
this strongly correlated region. 
On the other hand, from the result of renormalization group approach~\cite{rf:Fisher} for $t_3=0$ and 
$U\rightarrow 0^+$, 
the spin-singlet state is stable for any values of $t_2$ and $n$.
It does not contradict this result within the perturbation theory. 
When $U$ is very small, the vertex corrections are negligible which is 
important for the stabilization of the spin-triplet state. 
Therefore, when $U$ is very small, the spin-singlet state seems to be 
dominant from the results without the third-order terms of the pairing 
interaction. 
However, numerical calculation is difficult since the eigenvalue is very 
small in this case. 
\subsection{Momentum dependence of the anomalous self-energy}
\begin{figure}[t]
\begin{center}
\includegraphics[width=8cm]{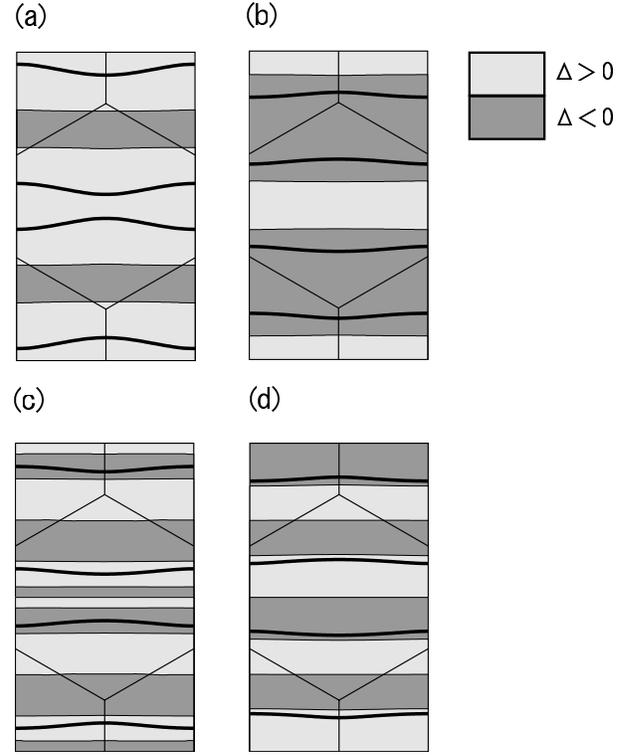}
\end{center}
\caption{(a),(b) Contour plots of the anomalous self-energy $\Delta(k)=0$ for the spin-singlet state in the case of $t_2=1.0$, $t_3=0.15$, $T=0.01$, $n=0.80$ 
and $U=4.0$. The thick lines represent the Fermi surfaces.
The spin-singlet state is a fully gapped state. 
(c),(d) Contour plots of the anomalous self-energy $\Delta(k)=0$ 
for the spin-triplet state in the case of $t_2=0.4$, $t_3=0.10$, $T=0.01$, 
$n=0.80$ and $U=4.0$. The thick lines represent the Fermi surfaces.
The spin-triplet state is a fully gapped state. }
\label{fig:gap}
\end{figure}
From eq. (\ref{eq:trans}), anomalous Green's functions on the two bands are 
given by 
\begin{equation}
\begin{split}
&F_{+}(k)=|G_{+}(k)|^2\Delta_{+}(k)=F_{\rm AA}(k)+F_{\rm AB}(k),\\
&F_{-}(k)=|G_{-}(k)|^2\Delta_{-}(k)=F_{\rm AA}(k)-F_{\rm AB}(k). 
\end{split}
\end{equation}
Therefore, anomalous self-energies on the two bands are given by 
\begin{equation}
\begin{split}
&\Delta_{+}(k)=\Delta_{\rm AA}(k)+\Delta_{\rm AB}(k),\\
&\Delta_{-}(k)=\Delta_{\rm AA}(k)-\Delta_{\rm AB}(k).
\end{split}
\end{equation}
In Fig. \ref{fig:gap}, we show contour plots of the anomalous 
self-energy in the case of $T=0.01$, $n=0.80$ and $U=4.0$. 
In Fig. \ref{fig:gap} (a) and (b), a spin-singlet case for 
$t_2=1.0$, $t_3=0.15$, 
and in Fig. \ref{fig:gap} (c) and (d), a spin-triplet case for
$t_2=0.40$, $t_3=0.10$ are shown. 
The thick lines represent the Fermi surfaces. 
For the spin-singlet state, the momentum dependence of the anomalous 
self-energy on the Fermi surface is a fully gapped state, 
and the signs of the anomalous self-energy on the Fermi surface for the two 
band are different. We discuss this result later in Sec. \ref{section:dis}. 
For the spin-triplet state, the momentum dependence of 
the anomalous self-energy is also a fully gapped state. 
\section{Discussions}
\label{section:dis}
In this section we consider the physical origin of the results in Sec. \ref{section:res}. 
Hereafter, we only consider the RPA-like terms to understand the 
mechanism of stabilization for the spin-singlet state. 
\'Eliashberg equation is written by 
\begin{equation}
\begin{split}
\Delta_{\rm AA}(k)=-\frac{T}{N}\sum_{k'}V_{\rm AA}(k,k')F_{\rm AA}(k'),\\
\Delta_{\rm AB}(k)=-\frac{T}{N}\sum_{k'}V_{\rm AB}(k,k')F_{\rm AB}(k').
\end{split}
\end{equation}
Here, we write $V^{\rm RPA\,Singlet}_{\rm AA,AA}(k,k')$ and 
$V^{\rm RPA\,Singlet}_{\rm AB,AB}$ as $V_{\rm AA}(k,k')$ and $V_{\rm AB}(k,k')$, respectively. 
The \'Eliashberg equation for $\Delta_{+}(k)$ and $\Delta_{-}(k)$ are 
given by 
\begin{equation}
\begin{split}
&\Delta_{+}(k)=\Delta_{\rm AA}(k)+\Delta_{\rm AB}(k)\\
&=-\frac{T}{N}\sum_{k'}\frac{1}{2}(V_{\rm AA}(k,k')+V_{\rm AB}(k,k'))
F_{+}(k')\\
&-\frac{T}{N}\sum_{k'}\frac{1}{2}(V_{\rm AA}(k,k')-V_{\rm AB}(k,k'))F_{-}(k')\\
&=-\frac{T}{N}\sum_{k'}\frac{1}{2}(V_{\rm AA}(k,k')+V_{\rm AB}(k,k'))
|G_{+}(k')|^2\Delta_{+}(k')\\
&-\frac{T}{N}\sum_{k'}\frac{1}{2}(V_{\rm AA}(k,k')-V_{\rm AB}(k,k'))
|G_{-}(k')|^2\Delta_{-}(k')\\
&=-\frac{T}{N}\sum_{k'}\frac{1}{2}V_{\rm intra}(k,k')
|G_{+}(k')|^2\Delta_{+}(k')\\
&-\frac{T}{N}\sum_{k'}\frac{1}{2}V_{\rm inter}(k,k')
|G_{-}(k')|^2\Delta_{-}(k'),
\end{split}
\end{equation}
\begin{equation}
\begin{split}
&\Delta_{-}(k)=\Delta_{\rm AA}(k)-\Delta_{\rm AB}(k)\\
&=-\frac{T}{N}\sum_{k'}\frac{1}{2}(V_{\rm AA}(k,k')-V_{\rm AB}(k,k'))F_{+}(k')\\
&-\frac{T}{N}\sum_{k'}\frac{1}{2}(V_{\rm AA}(k,k')+V_{\rm AB}(k,k'))F_{-}(k')\\
&=-\frac{T}{N}\sum_{k'}\frac{1}{2}(V_{\rm AA}(k,k')-V_{\rm AB}(k,k'))|G_{+}(k')|^2\Delta_{+}(k')\\
&-\frac{T}{N}\sum_{k'}\frac{1}{2}(V_{\rm AA}(k,k')+V_{\rm AB}(k,k'))|G_{-}(k')|^2\Delta_{-}(k')\\
&=-\frac{T}{N}\sum_{k'}\frac{1}{2}V_{\rm inter}(k,k')
|G_{+}(k')|^2\Delta_{+}(k')\\
&-\frac{T}{N}\sum_{k'}\frac{1}{2}V_{\rm intra}(k,k')
|G_{-}(k')|^2\Delta_{-}(k'),
\end{split}
\end{equation}
where, 
\begin{equation}
\begin{split}
&V_{\rm intra}(k,k')\equiv V_{\rm AA}(k,k')+V_{\rm AB}(k,k')\\
&V_{\rm inter}(k,k')\equiv V_{\rm AA}(k,k')-V_{\rm AB}(k,k').
\end{split}
\end{equation}
The pairing interaction which connects the anomalous self-energies on 
the intra band is 
$V_{\rm intra}(k,k')\equiv V_{\rm AA}(k,k')+V_{\rm AB}(k,k')$ 
and that on the inter 
band is $V_{\rm inter}(k,k')\equiv V_{\rm AA}(k,k')-V_{\rm AB}(k,k')$. 
Here, 
\begin{equation}
\begin{split}
&V_{\rm intra}(k,k')=V_{\rm AA}(k;k')+V_{\rm AB}(k;k')\\
&=U+U^2(\chi^{(0)}_{\rm AA}(k-k')+\chi^{(0)}_{\rm AB}(k-k'))\\
&+2U^3(\chi^{(0)}_{\rm AA}(k-k')+\chi^{(0)}_{\rm AB}(k-k'))^2\\
&=U+U^2\chi^{(0)}_{\rm intra}(k-k')+2U^3\chi^{(0)}_{\rm intra}(k-k')^2,\\
&V_{\rm inter}(k,k')=V_{\rm AA}(k,k')-V_{\rm AB}(k;k')\\
&=U+U^2(\chi^{(0)}_{\rm AA}(k-k')-\chi^{(0)}_{\rm AB}(k-k'))\\
&+2U^3(\chi^{(0)}_{\rm AA}(k-k')-\chi^{(0)}_{\rm AB}(k-k'))^2\\
&=U+U^2\chi^{(0)}_{\rm inter}(k-k')+2U^3\chi^{(0)}_{\rm inter}(k-k')^2, 
\end{split}
\end{equation}
where, 
\begin{equation}
\begin{split}
&\chi_{\rm intra}^{(0)}(q)\equiv\chi^{(0)}_{\rm AA}(q)+\chi^{(0)}_{\rm AB}(q),\\
&\chi_{\rm inter}^{(0)}(q)\equiv\chi^{(0)}_{\rm AA}(q)-\chi^{(0)}_{\rm AB}(q).
\end{split}
\end{equation}
Therefore,  $V_{\rm intra}(k,k')$ and $V_{\rm inter}(k,k')$ are composed of $\chi_{\rm intra}^{(0)}(k-k')\equiv\chi^{(0)}_{\rm AA}(k-k')+\chi^{(0)}_{\rm AB}(k-k')$ 
and $\chi_{\rm inter}^{(0)}(k-k')\equiv\chi^{(0)}_{\rm AA}(k-k')-\chi^{(0)}_{\rm AB}(k-k')$, respectively. 
Here, 
\begin{equation}
\begin{split}
&\chi_{\rm intra}^{(0)}(q)=\chi_{\rm AA}^{(0)}(q)+\chi_{\rm AB}^{(0)}(q)\\
&=-\frac{T}{N}\sum_{k'}\frac{1}{2}(G_{+}^{(0)}(k')G_{+}^{(0)}(q+k')+G_{-}^{(0)}(k')G_{-}^{(0)}(q+k')),\\
&\chi_{\rm inter}^{(0)}(q)=\chi_{\rm AA}^{(0)}(q)-\chi_{\rm AB}^{(0)}(q)\\
&=-\frac{T}{N}\sum_{k'}(G_{+}^{(0)}(k')G_{-}^{(0)}(q+k')).
\end{split}
\end{equation}
Therefore, $\chi_{\rm intra}^{(0)}(q)$ is the 
average of the bare susceptibility of the intra bands and
$\chi_{\rm inter}^{(0)}(q)$ is 
the bare susceptibility of the inter bands. 
\begin{figure}[t]
\begin{center} 
\includegraphics[width=8cm]{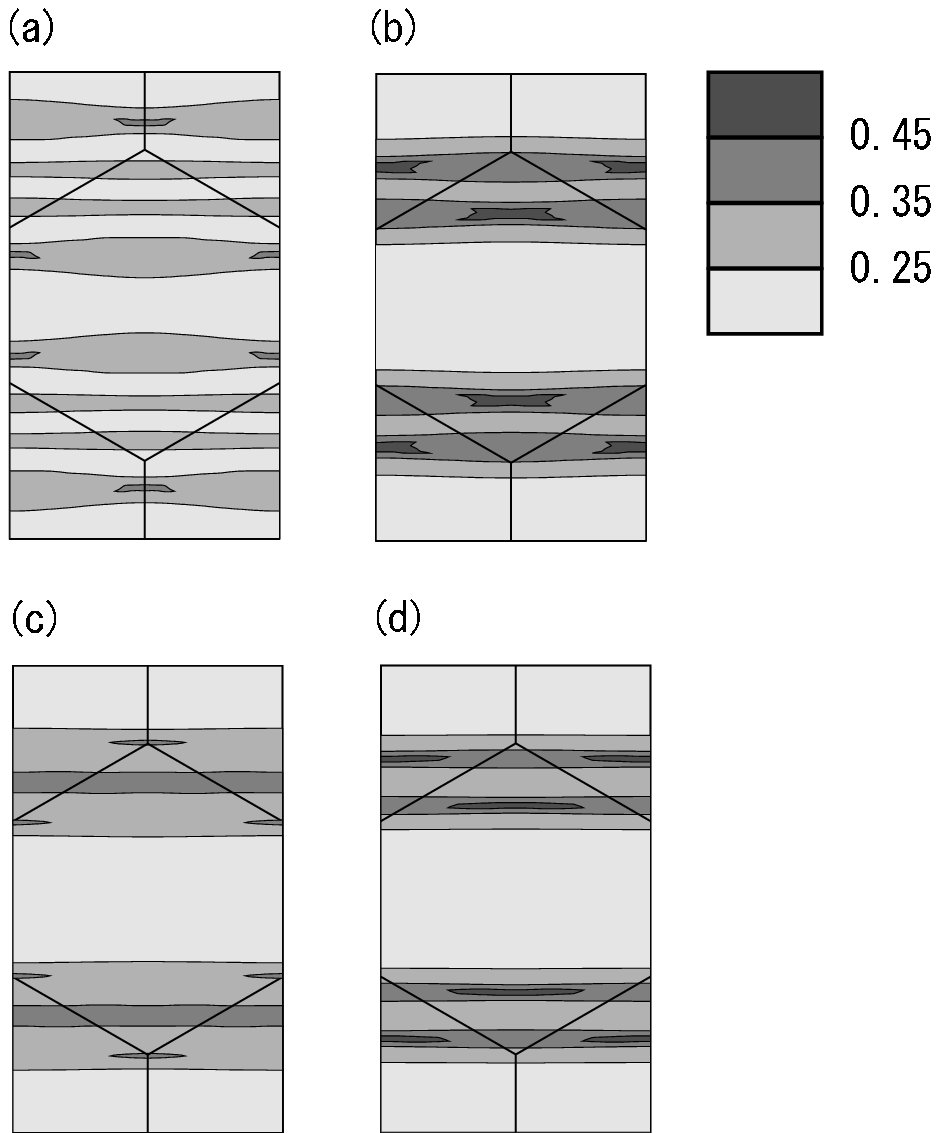}
\end{center}
\caption{
(a) A contour plot of 
$\chi_{\rm intra}^{(0)}(\ve{q},0)\equiv\chi_{\rm AA}^{(0)}(\ve{q},0)+\chi_{\rm AB}^{(0)}(\ve{q},0)$ 
in the case of $t_2=1.0$, $t_3=0.15$, $n=0.80$ and $T=0.01$.
(b) A contour plot of 
$\chi_{\rm inter}^{(0)}(\ve{q},0)\equiv\chi_{\rm AA}^{(0)}(\ve{q},0)-\chi_{\rm AB}^{(0)}(\ve{q},0)$ 
in the case of $t_2=1.0$, $t_3=0.15$, $n=0.80$ and $T=0.01$.
(c) A contour plot of 
$\chi_{\rm intra}^{(0)}(\ve{q},0)\equiv\chi_{\rm AA}^{(0)}(\ve{q},0)+\chi_{\rm AB}^{(0)}(\ve{q},0)$ 
in the case of $t_2=0.40$, $t_3=0.10$, $n=0.80$ and $T=0.01$.  
(d) A contour plot of 
$\chi_{\rm inter}^{(0)}(\ve{q},0)\equiv\chi_{\rm AA}^{(0)}(\ve{q},0)-\chi_{\rm AB}^{(0)}(\ve{q},0)$ 
in the case of $t_2=0.40$, $t_3=0.10$, $n=0.80$ and $T=0.01$. 
When the band gap between two band is large, 
$\chi_{\rm intra}^{(0)}(\ve{q},0)$ is small 
like the case (a). On the other hand, when the band gap between two band is 
small, $\chi_{\rm intra}^{(0)}(\ve{q},0)$ is large 
like the case (c). The spin-singlet state is stable in the case (a), 
and the spin-singlet state is unstable in the case (c). }
\label{fig:chi}
\end{figure}

Fig. \ref{fig:chi} is contour plots of 
$\chi_{\rm intra}^{(0)}(\ve{q},0)$ 
and $\chi_{\rm inter}^{(0)}(\ve{q},0)$ in the cases of 
$t_2=1.0$, $t_3=0.15$, $n=0.80$, $T=0.01$ and for $t_2=0.40$, $t_3=0.10$, 
$n=0.80$, $T=0.01$, respectively. 
Since the Fermi surfaces have nesting property, they have quasi-one 
dimensional peaks. 
Since $\chi_{\rm intra}^{(0)}(\ve{q},0)$ 
is the average of the bare susceptibility of the intra bands, 
it is usually 
smaller than $\chi_{\rm inter}^{(0)}(\ve{q},0)$. 
But when the band gap between two bands is small, 
two nesting vectors of intra band become almost same. 
Therefore, $\chi_{\rm intra}^{(0)}(\ve{q},0)$ with small band gap is 
large compared with the case where the band gap between two bands is large. 
Actually, in Fig \ref{fig:chi}, 
when the band gap between two bands is large, 
$\chi_{\rm intra}^{(0)}(\ve{q},0)$ is small 
like the case (a). On the other hand, when the band gap between two bands is 
small, $\chi_{\rm intra}^{(0)}(\ve{q},0)$ is large 
like the case (c). 
By using the results in Fig. 10, we can understand the 
superconducting gap structure 
in Fig. \ref{fig:gap} 
(a),(b) and the suppression of $\lambda_{\rm max}$ in the case where the 
band gap between two bands is small, as we explain in the following. 
From the structure 
in the \'Eliashberg equation,  
when $\chi_{\rm intra}^{(0)}(\ve{q},0)$ is 
much smaller than 
$\chi_{\rm inter}^{(0)}(\ve{q},0)$, 
in order to obtain a positive value of $\lambda_{\rm max}$, it is favorable 
that signs of $\Delta(k)$ on one band are different from 
its sign on another band. 
The structure of $\Delta(k)$ in Fig. \ref{fig:gap} (a),(b) just becomes so. 
But, when the band gap between two bands is small, 
two nesting vectors of intra bands are almost same. 
Therefore, $\chi^{(0)}_{\rm intra}(\ve{q},0)$ 
becomes large and 
the spin-singlet state is suppressed by the conflict of the peaks of 
$\chi_{\rm intra}^{(0)}(\ve{q},0)$ and 
$\chi_{\rm inter}^{(0)}(\ve{q},0)$. 
Actually, in Fig. \ref{fig:chi}, 
when the band gap between two bands is large, 
$\chi_{\rm intra}^{(0)}(\ve{q},0)$ is small 
and the spin-singlet state is very stable like the case (a). 
On the other hand, when the band gap between two bands is 
small, $\chi_{\rm intra}^{(0)}(\ve{q},0)$ is large 
and the spin-singlet state is unstable like the case (c).

Finally, we discuss the pairing symmetry 
in Sr$_{14-x}$Ca$_x$Cu$_{24}$O$_{41}$. 
From this calculation, we can see that the spin-singlet and fully 
gapped state is very stable 
and is consistent with the calculation within FLEX~\cite{rf:Kontani}. 
On the other hand, when the electron number 
density is shifted from the half-filled state and the band gap between two 
bands is small, the spin-triplet and fully gapped state is stable. 
We cannot unfortunately determine the electronic structure under the pressure 
at present. 
In both cases, the fully gapped state is not contradict to the experiment of 
$1/T_1$. 
On the other hand, the Knight shift does not change above and below $\Tc$. 
It suggests that a spin-triplet state is realized. But, since the paramagnetic 
contribution of Knight shift should be small owing to the effect of spin gap 
conformed at rather high temperatures, it might be 
difficult to detect the shift at $\Tc$ within the experimental accuracy. 
Here, we discuss the ratio of hopping integrals. 
The ratio $t_1/t_2$ may be considered to be unity 
from almost equivalent spacings of the leg and the rung in this ladder. 
Experimentally, however, the ratio of the spin exchange coupling constant 
$J_{\rm leg}/J_{\rm rung}$ is about twice 
for related compounds.~\cite{rf:Johnston,rf:Imai}
Here, $J_{\rm leg}$ and $J_{\rm rung}$ are related to $t_1$ and $t_2$, 
respectively. 
Moreover, the pressure might change the ratio $t_1/t_2$. 
Therefore, we can expect a flexible value for the ratio $t_1/t_2$.
\section{Conclusion}
In conclusion, we have investigated the pairing symmetry and the transition 
temperature on the basis of the trellis-lattice Hubbard model. 
We have solved the \'Eliashberg equation using the third-order perturbation 
theory with respect to the on-site repulsion $U$. 
We find that the spin-singlet state is strongly stable in a wide range of 
parameters. 
On the other hand, when the electron number 
density is shifted from the half-filled state and the band gap between 
two bands is small, the spin-triplet state is expected. 
Thus, we suggest the possibility of unconventional superconductivity in 
Sr$_{14-x}$Ca$_x$Cu$_{24}$O$_{41}$. 

\section{Acknowledgments}

Numerical calculation in this work was carried out at 
the Yukawa Institute Computer Facility.



\begin{thebibliography}{99}
\bibitem{rf:comment}
From the tight-binding model, $V(k)$ is complex number, but phase factor can be included into the $A$ and $A^{\dagger}$
\end{thebibliography}
\end{document}